%% file: products-arxiv.tex
\documentclass[footinbib,superbib,eqsecnum,prb,nobibnotes,showpacs]{revtex4}

\usepackage{hyperref}
\usepackage{amsmath,amsthm,amssymb}
\usepackage[latin1]{inputenc}

\input products-macros.inc

\numberwithin{defn}{section}

\begin{document}

\title{Operator product expansions \\ as a consequence of phase space properties}
\author{Henning Bostelmann}
 \email[Electronic mail: ]{bostelm@theorie.physik.uni-goettingen.de}
\affiliation{%
Universität Göttingen, Institut für Theoretische Physik, 
37077 Göttingen, Germany
}
\date{August 7, 2005}


\begin{abstract}
The paper presents a model-independent, 
nonperturbative proof of operator product
expansions in quantum field theory. As an input, a recently proposed
phase space condition is used that allows a precise description
of point field structures. Based on the product expansions,
we also define and analyze normal products (in the sense of Zimmermann).

\end{abstract}

\maketitle

\section{Introduction}

Quantum fields, representing observables sharply localized in space-time,
generally are quite singular objects.\cite{FreHer:pointlike_fields}
In particular, their products at coinciding space-time points are ill-defined
and lead to divergences. Since such products play a vital rôle in the definition
and classification of models (by means of path integrals or field equations),
a thorough understanding and precise description of their singularities is
of considerable interest.

A step toward this goal was taken by Wilson, \cite{Wil:non-lagrangian} who proposed
that a product of two fields $\phi(x)$ and $\phi'(y)$ should be expandable into
a series,
\begin{equation} \label{WilsonExpansion}
\phi(x) \cdot \phi'(y) = \sum_j c_j(\xmy) \phi_j \big(\frac{x+y}{2}\big),
\end{equation}
where $\phi_j(\cdotarg)$ are local fields as well, and $c_j(\cdotarg)$ are
(generalized) functions which show singularities at the origin. This
\emph{operator product expansion} should then be valid at short
distances, i.e., in the limit $x \to y$.

The operator product expansion provides a detailed description of the product's
singular behavior. It may furthermore serve to define ``normal products'' of fields,
in generalization of the Wick product. Any field $\phi_j$ in Eq.~\eqref{WilsonExpansion}
whose coefficient $c_j(\cdotarg)$ does not vanish in the limit may be taken
as a candidate for such a normal product. \cite{Zim:Brandeis}

However, it remains to clear up in which sense and under which conditions
Eq.~\eqref{WilsonExpansion} holds precisely. The expansion has originally been
proposed in perturbation theory \cite{Wil:non-lagrangian,Zim:Brandeis}
and has widely been used as a tool in Lagrangian field theory.\cite{LagrangianTool}
Investigations in the framework of axiomatic field theory
\cite{AxiomaticProducts} aimed at a rigorous proof;
it seems that the Wightman axioms are too weak to ensure existence of the expansion structure,
and that additional conditions are needed. Unfortunately, these conditions could
not be connected directly to physical properties of the theory.
More detailed results are available in conformal field theory, \cite{ConformalOPE}
especially in $1+1$-dimensional models. \cite{TwodConformalOPE}
Here the expansion is a consequence of conformal symmetry. These methods cannot be carried
over to physically more realistic situations, though.

This paper presents an approach that explains operator product
expansions in a model-independent way, based on physically motivated
assumptions. We make use of the theory's phase space behavior,
which has recently been shown to have a strong impact on 
the field content.\cite{HaaOji:germs,Bos:fields-article}
A phase space condition proposed in Ref.~\onlinecite{Bos:fields-article} can be taken as
a physically natural assumption to ensure a regular
short distance behavior. On these grounds, we will establish
operator product expansions rigorously in the sense of an asymptotic series.
Furthermore, we will define normal products of fields and investigate
their properties.

In Sec.~\ref{pfsect}, we recall the relevant facts and results on
the point field structure established in Ref.~\onlinecite{Bos:fields-article},
which lie at the root of our investigation. Section~\ref{prodsect}
defines products of pointlike fields 
and gives a proof of their product expansion. Normal products and their
properties are discussed in Sec.~\ref{npsect}. We end with a brief
outlook on the classification of models in Sec.~\ref{conclsect}.

The present work is based on the author's thesis; \cite{Bos:Operatorprodukte}
it presents an abbreviated and partially improved version of material
developed there. The reader is referred to the original text for a more
detailed exposition, especially regarding mathematical aspects and
the development of proofs.

\section{Pointlike fields and phase space properties}  \label{pfsect}

It has been outlined in 
Refs.~\onlinecite{HaaOji:germs} and \onlinecite{Haa:models} that a
strong connection exists between the phase space properties of a quantum field theory
and its point field structure. A natural phase space condition, similar
but not identical to the usual compactness or nuclearity conditions,
\cite{Nuclearity}
was proposed in Ref.~\onlinecite{Bos:fields-article} and shown to allow a precise
description of the theory's field content. We will give a brief review
of the results established in Ref.~\onlinecite{Bos:fields-article}, mainly to fix
our notation.

We start from a quantum field theory in the framework of Local Quantum Physics
\cite{Haa:LQP} in the vacuum sector, i.e., given by a net of observable algebras
$\ocal \mapsto \afk(\ocal)$, where $\ocal \subset \mcal := \rbb^{s+1}$ are open
bounded regions in Minkowski space of $s+1$ space-time dimensions, 
and $\afk(\ocal)$ are $W^\ast$-algebras
acting on some common Hilbert space $\hcal$. We assume the standard axioms:
isotony, locality, covariance with respect to some 
strongly continuous unitary representation $U(x,\Lambda)$
of the connected Poincaré group $\pfk_+^\uparrow$, and the spectrum condition
(positivity of energy). We will denote the elements of the translation subgroup 
of $\pfk_+^\uparrow$ as
$U(x) = \exp (i P_\mu x^\mu)$, where $H := P_0$ is the Hamiltonian;
its spectral projections will be written as $P_H(E)$. 
$\Sigma$ stands for the predual space of $\boundedops$,
i.e. the space of weak-$\ast$-continuous linear functionals on $\boundedops$,
the positive normed elements of which represent the physical states of the system.

As we restrict our attention to point fields fulfilling polynomial $H$-bounds
(cf. Ref.~\onlinecite{FreHer:pointlike_fields}), we consider the space
\begin{equation}
 \cinftys := \bigcap_{\ell>0} R^\ell \Sigma R^\ell , \qquad
 \text{where} \; R = (1+H)^{-1},
\end{equation}
equipped with the topology of simultaneous convergence in the norms
$\lnorm{\sigma}{\ell} := \|R^{- \ell} \sigma R^{- \ell}\|$, $\ell>0$. We also make use of its
dual space $\cinftys^\ast$, equipped with the weak topology; its elements
are linear forms $\phi$ which fulfill
\begin{equation}
  \lnorm{\phi}{\ell} := \| R^\ell \phi R^\ell \| < \infty
  \quad \text{for some $\ell>0$}.
\end{equation}
For linear maps $\alpha: \cinftys^\ast \to \cinftys^\ast$, we sometimes consider
\begin{equation}
  \lnorm{\alpha}{\ell,\ell'} = \| R^{\ell'} \alpha (R^{- \ell} \cdotarg R^{- \ell}) R^{\ell'} \|
  \qquad (\ell,\ell'>0)
\end{equation}
in case this expression is finite.

Let $\Psi$ be the space of linear continuous maps from $\cinftys$ to $\Sigma$,
where $\Sigma$ is equipped with the norm topology. 
For our setup, the inclusion map $\Xi \in \Psi$ plays a central rôle,
\begin{equation}
  \Xi : \quad  \cinftys \to \Sigma,  \quad \sigma \mapsto \sigma .
\end{equation}
In order to analyze this map in the short distance limit, we refer to the algebras
$\afk(r) := \afk(\ocal_r)$, where $\ocal_r$ is the standard double cone
of radius $r>0$ centered at the origin, then define for $\psi \in \Psi$,
\begin{equation}
  \lnorm{\psi}{\ell} _r := \lnorm{ \psi \restrict \afk(r) }{\ell}
  = \sup_{\sigma \in \Sigma} \sup_{A \in \afk(r)}
    \frac{ |\psi(R^\ell \sigma R^\ell) (A) | }{  \|\sigma\| \, \|A\| },
\end{equation}
which is finite for sufficiently large $\ell>0$, and for $\gamma \geq 0$ consider the pseudometrics
\begin{equation} \label{deltadef}
  \delta_\gamma (\psi) :=
  \begin{cases}
    0 & \text{if } \; r^{-\gamma} \lnorm{\psi}{\ell} _r \xrightarrow[r \to 0]{} 0
    \text{ for some $\ell>0$},  \\
    1 & \text{otherwise.}
  \end{cases}
\end{equation}
We say that the net $\afk$ fulfills the {\em microscopic phase space condition}
if for every $\gamma \geq 0$, there exists a map $\psi \in \Psi$ of finite
rank such that
\begin{equation}
 \delta_\gamma( \Xi-\psi ) = 0.
\end{equation}
It is not overly difficult to see that 
this condition holds for a wide range of free theories (see the Appendix of
Ref.~\onlinecite{Bos:fields-article}), and it seems plausible that 
the same condition is fulfilled in any model with
a sufficiently regular short distance behavior. Its consequences are the following:
There exists an increasing sequence of finite-dimensional vector spaces
$\Phi_\gamma \subset \cinftys^\ast$, $\gamma \geq 0$, the elements of which are Wightman fields
located at $x=0$.
Their union $\bigcup_\gamma \Phi_\gamma =: \PhiFH$ exhausts the field content
of the theory as investigated by Fredenhagen and Hertel. \cite{FreHer:pointlike_fields}
Given $\phi \in \PhiFH$, we can find a sequence $A_r \in \afk(r)$ ($r>0$) such that
\begin{equation} \label{fieldapprox}
  \lnorm{\phi- A_r}{\ell} = O(r), \quad
  \|A_r\| = O( r^{-k} )
\end{equation}
as $r \to 0$, where $k,\ell>0$ can be chosen uniformly for all $\phi \in \Phi_\gamma$,
with $\gamma$ being fixed. If $p_\gamma: \cinftys^\ast \to \cinftys^\ast$ is any continuous
projection
onto $\Phi_\gamma$, and $p_{\gamma \ast}$ its predual, then we find
\begin{equation}  \label{XiPgammaApprox}
  \delta_\gamma \big( \Xi - \Xi \circ p_{\gamma \ast} \big) = 0.
\end{equation}
[Note that a continuous projection can be found for any given 
finite-dimensional subspace $V \subset \cinftys^\ast$. 
We will require all such projections considered in the following
to be continuous and refrain from noting this fact repeatedly.]

Every $\Phi_\gamma$ is invariant under a certain class of symmetry transformations,
including Lorentz transformations, dilations and inner symmetries, 
provided that they exist as symmetries of the underlying net $\afk$.
Linear differential operators act as maps $\PhiFH \to \PhiFH$ as well, but typically
map $\Phi_\gamma$ into some larger space $\Phi_{\gamma'}$.

\section{Product expansions} \label{prodsect}

Our task will now be to establish operator product expansions, assuming
that the theory under discussion fulfills the microscopic phase space condition.
We will prove the expansion in the form
\begin{equation} \label{exp_toprove}
\phi(x) \cdot \phi'(y) = \sum_j c_j(x,y) \phi_j (0)
\qquad (x,y \to 0),
\end{equation}
which can easily be transformed into the more familiar form \eqref{WilsonExpansion}
and vice versa.

The proof is based on the following heuristic idea: Let $A,A' \in \afk(r)$ be
two {\em bounded} localized observables, and $A(x) := U(x) A U(x)^\ast$, etc.
Choosing a projector $p_\gamma$ onto $\Phi_\gamma$, 
the phase space property \eqref{XiPgammaApprox} can roughly be expressed as
$\Xi \approx \Xi \circ p_{\gamma\ast}$ at small distances. This means
\begin{equation}
  A(x) A'(y) \approx p_\gamma \big( A(x) A'(y) \big)
\end{equation}
if $x$, $y$ and $r$ are small enough. Expanding $p_\gamma = \sum_j \sigma_j(\cdotarg) \phi_j$
in a finite basis, this reads
\begin{equation}
  A(x) A'(y) \approx \sum_j \sigma_j \big( A(x) A'(y) \big) \phi_j
\end{equation}
in the limit $x,y,r \to 0$. If now $A \approx \phi$ and $A' \approx \phi'$ in this limit, 
we can define $c_j(x,y) := \sigma_j \big( A(x) A'(y) \big)$ 
and arrive at the expansion \eqref{exp_toprove}.

In order to transfer this heuristic idea into a rigorous expression for the product of
pointlike fields, we will first define the product of two fields 
[the left-hand side of Eq.~\eqref{exp_toprove}] 
for spacelike-separated $x$ and $y$, namely by means of
approximating them with bounded observables; this is done in Sec.~\ref{spprodsect}.
On these grounds, we will then establish the product expansion in Sec.~\ref{peproofsect}.
Some generalizations, such as products of more than two factors and products for
non-spacelike-separated arguments, will be discussed in Sec.~\ref{peextsect}.

\subsection{Spacelike products} \label{spprodsect}

In the following, let $\phi,\phi' \in \PhiFH$ be fixed. As a first step towards
the expansion \eqref{exp_toprove}, we will give meaning to the \emph{a priori}
ill-defined product $\phi(x) \cdot \phi'(y)$. This is certainly possible in a
distributive sense (as a product of smeared Wightman fields); however, we shall
use a more direct approach here. For $(x-y)^2 < 0$, we will define $\phi(x) \cdot \phi'(y)$
as an element of $\cinftys^\ast$, i.e., as a sesquilinear form.

This will be achieved by an approximation with bounded operators. We choose
sequences $A_r \to \phi$, $A'_r \to \phi'$ as specified in Eq.~\eqref{fieldapprox}.
Then it seems natural to define
\begin{equation} \label{heuristProdDefn}
 \phi(x) \cdot \phi'(y) := \lim_{r \to 0} A_r(x) A'_r(y),
 \qquad
 (x-y)^2 < 0.
\end{equation}
We will use methods from complex analysis to establish the existence of this limit
and to control the convergence dependent on $x$ and $y$.

To this end, let $\sigma = \hrskp{ \xi }{ \cdotarg | \xi' } $ be an energy-bounded
functional, i.e. $\xi^(\! \,' \!\,^) \in P_H(E) \hcal$ for some $E>0$. We assume
$\|\sigma\|=1$. Leaving $x$ fixed for the moment, we consider the function
\begin{equation}
  f_r(z) = \sigma \big( U(x) A_r U(z) A'_r U(-x-z) \big), \qquad
  z \in \mkr.
\end{equation}
$f$ is the boundary value of a function analytic on $\mkr + i \vlk$,
where $\vlk$ denotes the open forward light cone. This is seen from
\begin{equation} \label{frdef}
  f_r(z) = \sigma \big( U(x) A_r 
    \exp (i P_\mu z^\mu) A'_r U(x)^\ast \exp (-iP_\mu z^\mu) \big),
\end{equation}
using the spectrum condition and the strong continuity of translations.
[Note that the factor $\exp(-iP_\mu z^\mu)$ does not disturb analyticity here,
since $\sigma$ is energy-bounded.] We can estimate the modulus of $f_r$ on
$\mkr + i \vlk$ as
\begin{equation}
 | f_r(z) | \leq
  \|P_H(E) R^{- \ell}\| \, \|R^\ell A_r R^\ell \|  \,  
  \|R^{- \ell} \exp (i P_\mu z^\mu) R^{- \ell} \|  \,
% \\
% \times 
 \|R^\ell A'_r R^\ell \|  \,   \| R^{- \ell} P_H(E) \| \, \| P_H(E) \exp (-i P_\mu z^\mu)  \|.
\end{equation}
Here $\| R^\ell A_r \! \evp R^\ell \| $ 
stays bounded in the limit $r \to 0$ if $\ell$ is sufficiently large.
A straightforward calculation shows that
\begin{equation}
  \|R^{-2 \ell} \exp(i P_\mu z^\mu)\|  \leq  \| \im z \|^{-2 \ell} \cdot c_1,
  \quad
  \| P_H(E) \exp(-i P_\mu z^\mu)  \|  \leq e^{ c_2 E \|\im z\| }
  \quad \text{for}\; \|\im z\| \leq 1,
\end{equation}
where $c_1, c_2>0$ can be chosen constant if $\im z$ varies over some
open convex cone $\ccal$ with $\bar \ccal \subset \vlk$, which we keep 
fixed in the following.  (All $c_i$ will be positive constants in what follows.) 
Thus we have
\begin{equation}   \label{frbounds}
 | f_r(z) | \leq (1+E)^{2 \ell}\; e^{c_2 E \|\im z\|} \; \|\im z\|^{-2 \ell} \cdot c_3
% \\
\quad
 \text{for} \; r \leq 1, \; \im z \in \ccal, \; \|\im z \| \leq 1.
\end{equation}
By these and similar arguments, it follows 
in particular that $\{f_r\}_{r \leq 1}$ is a normal family of
analytic functions on $\mkr + i \vlk$, i.e., uniformly bounded on compact
subsets.

In just the same way, we may represent
\begin{equation} \label{frhatdef}
  \hat f_r(z) = \sigma \big( U(x+z) A'_r U(-z) A_r U(-x) \big)
\end{equation}
as a boundary value of functions analytic on $\mkr - i \vlk$; similar bounds as
in Eq.~\eqref{frbounds} can be established. If we choose $z$ real and spacelike, the
operators $A_r$ and $A'_r(z)$ will commute for small $r$, hence we find
\begin{equation} \label{frfrhatcoinc}
  f_r(z) = \hat f_r(z) \quad
  \text{for $z$ real, $z^2 < 0$, and small $r$.}
\end{equation}
This allows us to apply the edge of the wedge theorem to $f_r$ and $\hat f_r$:
They are parts of a single function $f_r$ analytic on $\mkr \pm i \vlk$ and on some
complex neighborhood of those real points where \eqref{frfrhatcoinc} holds.

Let us investigate this common analytic continuation more closely. 
We choose
some fixed $v \in \mkr$, $u \in \ccal$ with $v^2 < 0$, $\|u\|=1$. For $r$
small enough, the function
\begin{equation}
  f_{r,v,u}(t) = \begin{cases}
                  f_r(v+tu) & \text{if } \im t \geq 0, \\
                  \hat f_r(v+tu) & \text{if } \im t < 0
               \end{cases}
\end{equation}
is analytic on $\cbb \backslash \rbb$ and continuous on some real
neighborhood of $t=0$, which is, e.g., given by
\begin{equation}
  |t| < \half d(v),
  \qquad
  \text{where}\;
  d(v) := \min \{1,\dist(v,\partial \vlk \cup -\partial \vlk)\},
\end{equation}
assuming $r < \frac{1}{4}d(v)$. Applying Painlevé's theorem,\cite{Edgewedgeproof}
we see that
$f_{r,v,u}$ is indeed analytic on $\cbb \backslash \{ t \,|\, \im t=0, |t| \geq \half d(v)  \}$.
From Eq.~\eqref{frbounds}, we can derive the estimate
\begin{equation} \label{frtbounds}
 |f_{r,v,u}(t)| \leq c_4 (1\!+\!E)^{2 \ell} \, e^{c_2 E |\im t|} \, |\im t|^{-2 \ell}
 \quad \text{if} \; |t| < \half d(v),\;r  \leq 1.
\end{equation}
A Phragmén-Lindelöf-type of argument (cf. Lemma 5.7 of Ref.~\onlinecite{Bos:Operatorprodukte})
then leads us to 
\begin{equation} \label{plresult}
  |f_{r,v,u}(t) | \leq c_5 (1\!+\!E)^{2 \ell} e ^{\lambda c_2 E} \, \lambda^{-2 \ell}
  \quad
  \text{for $r < \frac{1}{4} d(v)$, $|t|<\lambda$ and any $\lambda \leq \frac{1}{4} d(v)$,}
\end{equation}
where $c_5$ depends neither on $v$ nor on $u$. Choosing $t$ to be purely imaginary,
this means
\begin{equation} \label{frboundsreal}
  |f_r(v+iu)| \leq c_5 (1+E)^{2 \ell} e^{\lambda c_2 E} \lambda^{-2 \ell} 
  %\\
  \quad
  \text{for}\;
  v^2 < 0, \, u \in \ccal, \, \|u\| < \lambda, \, \lambda \leq \frac{1}{4} d(v), \,
  r < \frac{1}{4} d(v).
\end{equation}
The same estimate holds at the real points ($u=0$), setting $t=0$ in Eq.~\eqref{plresult}.

As mentioned above, $f_r$ possesses an analytic continuation to a complex
neighborhood of any real spacelike $z$, provided that $r$ is sufficiently
small for $z$. This continuation is known very explicitly (see the proof of the
edge of the wedge theorem in Ref.~\onlinecite{Edgewedgeproof} for details); in fact,
the domain of holomorphy depends only on the geometry of $\ccal$,
and the estimate \eqref{frboundsreal} can be carried over to the continued
function, showing that $\{f_r\}_{r \leq 1}$ is a normal family throughout its domain.

We are now in the position to control the limit $r \to 0$. Since
$\|R^{-2 \ell} \exp(i P_\mu z^\mu) \| < \infty$ for $\im z \in \vlk$, the expression
\begin{equation} \label{proddef}
  f(z) := \sigma \big( U(x) \, \phi \,  U(z) \, \phi' \, U(-\!x\!-\!z) \big)
\end{equation}
is well-defined for $z \in \mkr + i \vlk$. Using the approximation properties
\eqref{fieldapprox} of $A_r$ and $A'_r$, and applying the same methods that
lead to Eq.~\eqref{frtbounds}, we can establish the estimate
\begin{multline} \label{fdiffbounds}
  |f(z)-f_r(z)| \leq c_6 (1\!+\!E)^{2 \ell} \; e^{ c_2 E |\im t |} \;  |\im t|^{-2 \ell} \cdot r \\
  \text{for}\;
  z = v + i t u, \, v \in \mkr, \, u \in \ccal, \, \|u\| = 1, \,
  |t| < \half d(v), \, r \leq 1.
\end{multline}
Combined with similar estimates for $\|\im z\| >1$, 
we see in particular that $f_r(z) \to f(z)$ for $z \in \mkr + i \vlk$.
In analogy to \eqref{frhatdef}, we may define $f(z)$ on $\mkr - i \vlk$;
pointwise convergence $f_r \to f$ and estimates of the form \eqref{fdiffbounds}
hold there, too. Now, since $\{f_r\}_{r \leq 1}$ is a normal family, we know that
$f_r$ converges to some analytic limit throughout its domain, 
where the convergence is uniform on compact subsets; 
so $f$ has an analytic continuation to that region,
and $f_r \to f$ holds in particular at the real points. Thus the limit
\begin{equation} \label{prodformdef}
  \sigma\big( \phi(x)  \cdot \phi'(y) \big) := \lim_{r \to 0} f_r(y-x)
\end{equation}
is well-defined and -- cf. Eq.~\eqref{proddef} -- independent of the choice
of sequences $A_r, A'_r$. This definition obviously is linear in $\sigma$;
so $\phi(x) \cdot \phi'(y)$ is a well-defined linear form on $P_H(E) \Sigma P_H(E)$
for any $E$.
In view of Eq.~\eqref{frboundsreal}, the following estimate holds:
\begin{equation}
  \|P_H(E) \, \phi(x) \cdot \phi'(y) \, P_H(E) \|
  \leq c_5 (1\!+\!E)^{2 \ell} e^{\lambda c_2 E} \lambda^{-2 \ell}
%  \\
\quad
  \text{for}\; (x-y)^2 < 0, \; \lambda \leq \frac{1}{4} d(\xmy).
\end{equation}
Applying this result with $\lambda = (c_2 E)^{-1}$ for $(c_2 E)^{-1} < \frac{1}{4} d(\xmy)$,
and with $\lambda = \frac{1}{4} d(\xmy)$ otherwise, we obtain a $c_7>0$ such that
\begin{equation} \label{prodEbounds}
  \|P_H(E) \, \phi(x) \cdot \phi'(y)  \, P_H(E) \|
  \leq c_7 (1\!+\!E)^{4 \ell} d(\xmy)^{-2 \ell}
  \quad
  \text{for any $E>0$.}
\end{equation}
Using the spectral representation of $R$, one can show on these grounds that
$\phi(x) \cdot \phi'(y) \in \cinftys^\ast$, and
\begin{equation}
 \lnorm{ \phi(x) \cdot \phi'(y) }{ 4 \ell+1 } \leq c_8 \, d(\xmy)^{-2 \ell}.
\end{equation}
The same methods can be applied to quantify the convergence $A_r(x) A'_r(y) \to \phi(x) \cdot \phi'(y)$.
Starting from Eq.~\eqref{fdiffbounds} and using the arguments that lead
to Eqs.~\eqref{frboundsreal} and \eqref{prodEbounds}, we arrive at the estimate
\begin{equation}
 \lnorm{ \phi(x) \cdot \phi'(y)  - A_r(x) A'_r(y) }{ 4 \ell+1 } \leq c_9 \, d(\xmy)^{-2 \ell} \, r
 \quad
 \text{for } r < \frac{1}{4} d(\xmy).
\end{equation}
To summarize, we have established the following result:

\begin{thm}   \label{spprodthm}
For $\phi,\phi' \in \PhiFH$, there exist linear forms
\begin{equation*}
  \phi(x) \cdot \phi'(y) \in \cinftys^\ast \qquad
  \text{for} \;
  x,y \in \mkr, \;\; (x-y)^2 < 0,
\end{equation*}
with the following properties: Given $\gamma$, we can choose constants
$\ell,m > 0$ and for any fixed $\phi,\phi' \in \Phi_\gamma$
another constant $c>0$ such that
\begin{equation*}
  \lnorm{ \phi(x) \cdot \phi'(y) }{m} \leq c \; d(\xmy)^{-2 \ell}.
\end{equation*}
If $A_r,A'_r \in \afk(r)$ are sequences of operators with
$  \lnorm{ \phi \evp - {A_r}\! \evp }{\ell} = O(r)$, then
\begin{equation*}
  \lnorm{ \phi(x) \cdot \phi'(y) - A_r(x) A_r'(y) }{m}
  \leq O(r)  \, d(\xmy)^{-2 \ell},
\end{equation*}
where the estimate $O(r)$ is uniform in $x$, $y$, given that $r < \frac{1}{4} d(\xmy)$.
\end{thm}

Note that the spacelike product usually diverges as $x \to y$,
say, on a straight line, but that the singularity is bounded by an inverse
power of $d(\xmy)$.

\subsection{Proof of the expansion} \label{peproofsect}

Based on the detailed results established for the spacelike product
$\phi(x) \cdot \phi'(y)$, we are now going to prove the operator product
expansion for this case, making use of the phase space approximation
\begin{equation} \label{phspapprox}
  \delta_\gamma \big( \Xi - \Xi \circ p_{\gamma \ast} \big) = 0.
\end{equation}
We follow the heuristic motivation given at the beginning of this section.
To be precise, let $\gamma \geq 0$ be given, and let $\phi,\phi' \in \Phi_\gamma$.
As before, we choose approximating sequences $A_r, A'_r \in \afk(r)$ with
\begin{equation} \label{fieldapprox2}
  \lnorm{\phi \evp - A_r \! \evp}{\ell} = O(r), \quad
  \|A_r \! \evp \| = O( r^{-k} ),
\end{equation}
where $k,\ell>0$ are suitably chosen; note that we can choose them dependent on $\gamma$ only.

First of all, we will approximate the bounded product $A_r(x) A'_r(y)$.
Let $\gamma' > 0$ (its value will be specified later), and fix a projector
$p_{\gamma'}$ onto $\Phi_{\gamma'}$.
Since $A_r(x) A'_r(y) \in \afk(r + 2 \|x\| + 2 \|y\|)$, it follows from
Eqs. \eqref{phspapprox} and \eqref{fieldapprox2} that for some $m>0$ (dependent on $\gamma$),
\begin{equation} \label{boundedprodapprox}
  \lnorm{ p_{\gamma'} \big( A_r(x) A'_r(y) \big) - A_r(x) A'_r(y) }{m}
  \leq (r + 2 \|x\| + 2 \|y\|)^{\gamma'} O(r^{-2k}).
\end{equation}
Since $\Phi_{\gamma'}$ is finite dimensional, we can find $m'\geq m$ such 
that $\lnorm{p_{\gamma'}}{m,m'} < \infty$; in fact, this choice depends on $\gamma'$ only.
Now on the left-hand side of Eq.~\eqref{boundedprodapprox}, 
$A_r(x) A'_r(y)$ converges to $\phi(x)\cdot \phi'(y)$ in the limit $r \to 0$:
In view of Theorem~\ref{spprodthm}, one sees that
\begin{equation} \label{fieldprodapprox}
  \lnorm{ p_{\gamma'} \big(\phi(x) \cdot \phi'(y) \big) - \phi(x) \cdot \phi'(y) }{m'}
%  \\
\;
  \leq (r + 2 \|x\| + 2 \|y\|)^{\gamma'}\, O(r^{-2k})
  + d(\xmy)^{-2 \ell} \,O(r),
\end{equation}
given that $r < \frac{1}{4} d(x-y)$. 

We will now consider the limit $x,y \to 0$, where we assume that
\begin{equation} \label{splimcond}
  d(\xmy) \geq (\|x\|+\|y\|) \cdot \mathrm{const},
\end{equation}
i.e., we demand that $\xmy$ does not approach the light cone too fast. We will
refer to this approximation as the {\em spacelike limit} and denote it by
$\splimit$. Now given some $\beta > 0$, we set
\begin{equation}
 r(x,y) := (\|x\|+\|y\|)^{1 + \beta + 2 \ell},
\end{equation}
which fulfills $r < \frac{1}{4} d(x-y)$ 
for small $x,y$ due to Eq.~\eqref{splimcond}. If now $\gamma'$
was chosen sufficiently large (dependent
on $\beta$), we see that \eqref{fieldprodapprox}
vanishes faster than $(\|x\|+\|y\|)^\beta$ in the spacelike limit. Thus, we have
achieved the following.

\begin{thm} \label{pe2thm}
Let $\gamma > 0, \beta > 0$ be given. 
We can find constants $\gamma'>0$ and $\ell>0$ 
such that for any $\phi,\phi' \in \Phi_\gamma$ and any projection $p_{\gamma'}$
onto $\Phi_{\gamma'}$:
\begin{equation*}
  (\|x\|+\|y\|)^{-\beta} \; \biglnorm{ \phi(x) \cdot \phi'(y) - p_{\gamma'} \big( \phi(x) \cdot \phi'(y) \big) }{\ell}
  \splimit 0.
\end{equation*}
\end{thm}

This establishes the operator product expansion -- as explained at the beginning of 
this section, we may expand $p_{\gamma'}$ in a basis in order to pass to the more
explicit form \eqref{exp_toprove}. In mathematical terms, product expansions
are asymptotic series;
while with increasing $\beta$ we will usually have to increase $\gamma'$,
to any finite approximation order $\beta$ a finite ``number of approximation terms''
will suffice. Note that the approximation terms are not unique (since $p_{\gamma'}$ is not);
however, the ambiguities are restricted to terms that vanish rapidly in the limit,
so at least the singular structure of an operator product expansion can be understood
as an intrinsic property of the theory. It is also worth noting that the approximation
is not only valid in the weak sense, as originally proposed in Ref.~\onlinecite{Wil:non-lagrangian},
but holds uniformly for all states of sufficiently regular high energy behavior.

Furthermore, the coefficients of the product expansion are simply matrix elements
of the spacelike product $\phi(x) \cdot \phi'(y)$; thus, we can apply
Theorem~\ref{spprodthm} directly, which shows that their divergences at
$x=y=0$ are bounded by an inverse power of $d(x-y)$.

\subsection{Further directions}                    \label{peextsect}

The results established in Secs.~\ref{spprodsect} and~\ref{peproofsect}
can be generalized in many ways. For the sake of brevity,
we will just sketch these findings; the reader is referred
to Chap.~5 of Ref.~\onlinecite{Bos:Operatorprodukte} for details of the construction.

First, we may consider products of arbitrary many factors;
though we restricted ourselves to the case of two factors in the above, 
our methods carry over quite directly.
In the following, let $n \in \nbb$ and $\gamma \geq 0$ be given. It seems natural
not only to consider $n$-fold products, but also their linear combinations;
by the same methods as outlined in Sec.~\ref{spprodsect}, we may define spacelike products
\begin{equation} \label{spprodmultfact}
 \Pi(x) = \sum_k c_k \phi_k^{(1)} (x^{(1)}) \cdot \ldots \cdot \phi_k^{(n)} (x^{(n)})
 \in \cinftys^\ast, \quad
   \text{where}\; c_k \in \cbb, \; \phi_k^{(j)} \in \Phi_\gamma.
\end{equation}
This expression is multilinear in the fields $\phi_k^{(j)}$ 
[note that the products $\phi(x)\cdot\phi'(y)$ were bilinear in $\phi,\phi'$
by definition],
so we can formally obtain the products 
from a map $\Pi \mapsto \Pi(x)$ which is well-defined on 
the tensor product space $\Phi_\gamma^{\otimes n}$.
In Eq.~\eqref{spprodmultfact}, we must demand that the components
of $x = (x^{(1)},\ldots,x^{(n)}) \in \mkr^n$ are (pairwise) spacelike separated. More abstractly, let
\begin{equation} \label{spsetdef}
 \mkr^n_{\mathrm{sp}} := \big\{ x \in \mkr^n \; \big| \; (x^{(i)}-x^{(j)})^2 < 0 \; \;\forall \, 1 \leq i < j \leq n     \big\},
\end{equation}
and define
\begin{equation}
 d(x) := \min \big\{1, \dist (x,\partial  \mkr^n_{\mathrm{sp}} )  \big\};
\end{equation}
we then demand of the spacelike limit $x \splimit 0$ that
\begin{equation}
  x \in  \mkr^n_{\mathrm{sp}}, \qquad
  \|x\| \leq d(x) \cdot \mathrm{const}.,
  \qquad
  x \to 0,
\end{equation}
where $\|x\|$ stands for the Euclidean norm of $x$ in $\mkr^n=\rbb^{n(s+1)}$.
Following the line of arguments given in Sec.~\ref{spprodsect}, 
the divergence of the spacelike products can be estimated as
\begin{equation}
  \lnorm{ \Pi(x) }{n \cdot m} = O \big( d(x)^{- n \cdot \ell} \big)
  \qquad
  (x \splimit 0),
\end{equation}
where $m, \ell$ depend on $\gamma$ only. Finally, we may obtain the following
analog of Theorem~\ref{pe2thm}.
\begin{thm} \label{pemultthm}
Let $\gamma > 0$, $\beta > 0$, $n \in \nbb$ be given.
We can find constants $\gamma'>0$ and $\ell>0 $ such that
for any $\Pi \in \Phi_\gamma^{\otimes n}$ and any projection $p_{\gamma'}$
onto $\Phi_{\gamma'}$,
\begin{equation*}
  \|x\|^{-\beta} \; \lnorm{ \Pi(x) - p_{\gamma'} \Pi(x) }{\ell}
  \splimit 0.
\end{equation*}
\end{thm}
Hence product expansions exist for products of an arbitrary (finite) number of fields.

Moreover, similar expansions can be established
for arbitrary (not necessarily spacelike) distances of arguments. Here the
field products are defined in the sense of distributions only, 
i.e., we replace $\Pi(x)$ with
$\Pi(f) \in \cinftys^\ast$, where $f \in \scal(\mkr^n)$ 
is a test function with compact support. 
Theorem~\ref{pemultthm} holds in an analogous way, where 
$\|x\|$  is substituted with 
\begin{equation}
   d(f) := \sup \{  \|x\|  \;|\; x \in \supp f \};
\end{equation}
instead of the spacelike limit, we consider the limit $d(f) \to 0$, and
we must require that for every multi-index $\mu$, a constant $c_\mu$
exists such that
\begin{equation}
	\| \partial^\mu f \|_{L_1} \leq d(f)^{-|\mu|} c_\mu
	\quad \text{ as }\; d(f) \to 0.
\end{equation}
So the product expansions can be extended to the non-space-like region, where
their coefficients are no longer functions, but rather
tempered distributions.

Furthermore, one may investigate the action of symmetry transformations
on the products and their expansions, demanding that these transformations
are compatible with translations and
with the product structure on $\boundedops$
(in a suitably defined way -- see Chap.~5.4 of Ref.~\onlinecite{Bos:Operatorprodukte}; these conditions are, e.g., fulfilled
for Lorentz transforms, dilations and inner symmetries).
The results are compatible with what is expected from perturbation theory.\cite{WeiBookOPE}

\section{Zimmermann's normal products}   \label{npsect}

In the preceding section, we have defined and analyzed products of fields at different
(spacelike separated) space-time points, and investigated 
their divergences at small distances.
However, with possible applications in mind, one would like to develop
some substitute for the ill-defined product at coinciding points in the sense
of a local field (i.e., an element of $\PhiFH$). In free field theory,
such a substitute is given by means of the Wick product, e.g., the normal ordered square
$\wickprod{\phi^2} (x)$ of a real scalar field. According to
Wick's theorem,\cite{WigGar:fields_as_distributions} it can be constructed
from spacelike products by subtraction of divergent terms,
\begin{equation} \label{wicksTheorem}
\wickprod{ \phi^2 } (0)
= \lim_{x,y \to 0}
\big(
\phi(x) \cdot \phi(y) -
 \hrskp{\Omega\,}{ \, \phi(x) \cdot \phi(y)  \,| \,\Omega}\, \idop
\big).
\end{equation}
In interacting theories, one would not expect such a limit to exist;
nevertheless, similar ``subtraction methods'' combined with a suitable
``renormalization factor'' can be used at least in a perturbative context
to justify the existence of local field equations.
\cite{PerturbFieldEq,Zim:field_equations_phi4}
Zimmermann \cite{Zim:Brandeis} used operator product expansions to derive
these constructions more generally: For any field $\phi_k$ occuring in the expansion~\eqref{WilsonExpansion},
one obtains heuristically
\begin{equation} \label{ZimmermannNP}
\phi_k(0) = \lim_{x,y \to 0} \frac{1}{c_k(\xmy)} \Big(
  \phi(x) \cdot \phi'(y) -
 \sum_{j \neq k} c_j(\xmy) \phi_j \big(\frac{x+y}{2} \big)
\Big),
\end{equation}
provided that the coefficient $c_k$ does not vanish in the limit. So every
``composite field operator'' $\phi_k$ that appears in the expansion
can serve as a candidate for a normal product.

Let us see how this can be formalized in our context. We fix a product
$\Pi \in \PhiFH^{\otimes n}$ and try to collect all ``relevant'' terms in its
product expansion. To this end, we make the following definition: We say that
a finite-dimensional subspace $V \subset \cinftys^\ast$ is {\em spacelike
approximating} for $\Pi$ if for some projection $p_V$ onto $V$ and some $\ell>0$,
\begin{equation} \label{spappdef}
  \lnorm{ \Pi(x) - p_V \Pi(x) }{\ell} \splimit 0.
\end{equation}
It can easily be seen, using the triangle inequality, that then the same is
true for {\em any} projection $p_V$ onto $V$. In view of Theorem~\ref{pemultthm},
$\Phi_\gamma$ always is spacelike approximating (sp-app) for $\Pi$ if $\gamma$ is chosen
sufficiently large. Moreover, a short calculation shows that if
$V$ and $W$ are two spaces which are sp-app for $\Pi$, 
then the same holds for $V \cap W$; this is easily extended to the intersection
$\bigcap_{i \in I} V_i$ of an arbitrary family $\{V_i\}_{i \in I}$ of 
sp-app spaces, even if $I$ is infinite.  
(Note that all spaces in question are finite dimensional.) 
That justifies the following definition.
\begin{defn}
Let $\Pi \in \PhiFH^{\otimes n}$. The 
finite-dimensional space
\begin{equation*}
  \npspace{\Pi} := \bigcap_{V \mathrm{sp-app~for~}\Pi} V \;\;\subset \PhiFH
\end{equation*}
is called the \emph{normal product space} of $\Pi$.
It is the smallest space that is spacelike approximating for $\Pi$.
\end{defn}
Let $p$ be a projection onto $\npspace{\Pi}$, and choose a basis
$\{\phi_j\}_{j=1}^J$ of $\npspace{\Pi}$. 
Since $\npspace{\Pi} \subset \PhiFH$, 
the basis elements $\phi_j$ are local Wightman fields.
Expanding $p$ in this basis, 
we find functions $c_j(x)$ such that
\begin{equation}
   p \Pi(x) = \sum_{j=1}^{J} c_j(x) \phi_j \,.
\end{equation}
Due to the minimality of $\npspace{\Pi}$, none of the coefficients $c_j(x)$
vanishes in the spacelike limit. Thus for every $k \in \{1,\ldots,J\}$, we can find a sequence
$(x_n)$ with $x_n \splimit 0$ such that
\begin{equation}
  \phi_k = \lim_{n \to \infty}
  \frac{1}{c_k(x_n)} \Big(
   \Pi(x_n) -
  \sum_{j \neq k} c_j(x_n) \phi_j    \Big)
\end{equation}
with respect to some norm $\lnorm{\cdotarg}{\ell}$; we have recovered
Zimmermann's approximation formula.

$\npspace{\Pi}$ is a normal product not in the sense of a single field,
but as a vector space containing all possible candidates for such a normal product field.
In the case of a real scalar free field $\phi(x)$, one obtains the
result (cf. Chap.~5.7 of Ref.~\onlinecite{Bos:Operatorprodukte})
\begin{equation}
  \npspace{ \phi \otimes \phi } = \lspan \{ \idop, \wickprod{\phi^2} \},
\end{equation}
so the normal product space gives us some generalization of the Wick product.
In free field theory, it is possible to choose a distinct element
$\wickprod{\phi^2 } \in \npspace{ \phi \otimes \phi}$ by virtue
of ``normal ordering'' or of Eq.~\eqref{wicksTheorem}.
This structure is lost in the general case, as is suggested by perturbation theory
and low-dimensional integrable models. Certainly, for specific applications, 
there may be additional restrictions on the choice of a normal product field.
One can try to isolate a ``most divergent term'' in the product expansion, 
seek for specific representations of the Lorentz group\cite{lorentzCurrents}
(see also below), or use field equations as selection criteria.\cite{Zim:field_equations_phi4}
Still, some ambiguities may remain;\cite{Joh:green_2d} 
in our general setting, it does not seem possible
to establish a full substitute for the Wick product.

We can slightly modify the methods developed above in order to define ``extended'' normal product spaces
$\npspace{\Pi}_\beta$ for $\beta \geq 0$, requiring that the left-hand side of
Eq.~\eqref{spappdef} vanishes faster than $\|x\|^\beta$ in the limit. That provides
us with an increasing sequence of vector spaces $\npspace{\Pi}_0 \subset \npspace{\Pi}_1 \subset \ldots$
containing higher and higher order composite field operators of some fixed
product $\Pi$. This construction has recently found application in a
characterization of nonequilibrum thermodynamical states. \cite{BOR:non-equilibrum}

We shall now investigate the behavior of $\npspace{\Pi}$ under Lorentz transformations
or other symmetries and under differential operators. All these cases will be treated
within a single concept. We consider a transformation $\alpha$ which acts in
three different ways (denoted by the same symbol for simplicity),

\begin{enumerate}
\renewcommand{\theenumi}{\roman{enumi}}
\renewcommand{\labelenumi}{(\theenumi)}
\item
a linear, continuous map $\alpha: \cinftys^\ast \to \cinftys^\ast$,

\item
linear maps $\alpha: \PhiFH^{\otimes n} \to \PhiFH^{\otimes n}$ (for every $n$),

\item
an invertible action $x \mapsto \alpha.x$ on $\mcal^n$ (for every $n$)

\end{enumerate}
with the following properties:
\begin{enumerate}
\renewcommand{\theenumi}{(\arabic{enumi})}
\renewcommand{\labelenumi}{\theenumi}
\item \label{compatibilityProp}
$ \quad \alpha(\Pi(x)) = (\alpha \Pi)(\alpha.x) \; \; 
\forall \, \Pi \in \PhiFH^{\otimes n}$, $n \in \nbb$, $x \in \mkr^n$,

\item \label{limitProp}
$ \quad \alpha .x \splimit 0 \equivalent x \splimit 0$,

\item \label{normProp}
$\quad \lnorm{ \alpha }{ \ell,\ell' } < \infty $ for any $\ell>0$ and appropriate $\ell'>0$ (dependent on $\ell$).

\end{enumerate}

We shall show that under these conditions, one has $\alpha \npspace{\Pi} = \npspace{\alpha \Pi}$;
applications of this ``covariance property'' will be discussed below.
As a first step, we shall prove the following lemma:

\begin{lemm} \label{projectionLemma}
  Let $V \subset \cinftys^\ast$ be a finite-dimensional subspace. There exist
  projections $p$ onto $V$ and $p'$ onto $\alpha V$ such that
  \begin{equation*}
    \alpha \circ p = p' \circ \alpha.
  \end{equation*}
\end{lemm}

\begin{proof}
Let $K:= \ker \alpha \cap V$. Choose a space $\hat V$ such that $V = K \oplus \hat V$.
Furthermore, choose projections $p_K$ onto $K$ such that $p_K\restrict \hat V = 0$,
and $p'$ onto $\alpha V = \alpha \hat V$. (This is certainly possible, since $V$ is finite-dimensional.) 
Denote the inverse of $\alpha: \hat V \to \alpha V$ by $\hat \alpha^{-1}$.
We define
\begin{equation}
  p := p_K + \hat \alpha ^{-1} \circ p' \circ \alpha.
\end{equation}
A short calculation shows $p^2=p$, $\img p = V$, $\alpha \circ p = p' \circ \alpha$,
so $p$ has the properties desired.
\end{proof}

Now we are in the position to prove the ``covariance'' of $\npspace{\Pi}$.

\begin{thm} \label{covarThm}
Let $\alpha$ fulfill the conditions \ref{compatibilityProp} to \ref{normProp}
listed above. Then
\begin{equation*}
   \npspace{\alpha \Pi}  = \alpha \npspace{\Pi} \quad
   \forall \; \Pi \in \PhiFH^{\otimes n}, \; n \in \nbb.
\end{equation*}
\end{thm}
\begin{proof}
Let $V \subset \cinftys^\ast$ be sp-app for $\Pi$, and $p,p'$
projections as in Lemma~\ref{projectionLemma}. Then for sufficiently large $\ell,\ell'$,
we have
\begin{equation}
  \lnorm{ (\alpha \Pi)(\alpha.x) - p' (\alpha \Pi) (\alpha.x) }{\ell}
  = \lnorm{ \alpha (\Pi(x) - p \Pi(x) )}{\ell}
%  \\
  \leq \lnorm{\alpha}{\ell,\ell'} \, \lnorm{\Pi(x) - p \Pi(x)}{\ell'} \splimit 0;
\end{equation}
due to property \ref{limitProp}, this means that
$\alpha V$ is sp-app for $\alpha \Pi$. Hence
$\npspace{\alpha \Pi} \subset \alpha \npspace{\Pi}$. 

To show the opposite
inclusion, split $V := \npspace{\Pi}$ into a direct sum
\begin{align}
&\npspace{\Pi} = V = V_0 \oplus V_1 \oplus V_2,
\\ \notag
\text{where}\quad &
V_0 = \ker \alpha \cap V, \quad
\alpha V_1 = \npspace{\alpha \Pi}, \quad
\alpha V_2 \cap \npspace{\alpha \Pi} = \{0\}.
\end{align}
Let $p_i : V \to V_i$ (i=0,1,2) denote the projection operators with respect
to that direct sum, and let $p_i': \alpha V \to \alpha V_i$ ($i=1,2$)
be the projections with regard to the direct sum $\alpha V = \alpha V_1 \oplus \alpha V_2$.
We then have $p_i' \alpha = \alpha p_i$ for $i=1,2$. 
We choose projections $p,p'$ as in Lemma~\ref{projectionLemma};
then $p_i \circ p: \cinftys \to V_i$ are projections onto $V_i$,
and $p_i' \circ p': \cinftys \to \alpha V_i$ are projections onto $\alpha V_i$.
Now, since both $\alpha V$
and $\alpha V_1 = \npspace{\alpha \Pi}$ are sp-app for $\alpha \Pi$,
we see that for sufficiently large~$\ell$,
\begin{equation}
\lnorm{ (\alpha \Pi)(x) - (p'_1+p'_2) p' (\alpha \Pi)(x) }{\ell} \splimit 0
\quad \text{and}
\quad
\lnorm{(\alpha \Pi)(x) -  p'_1 p' (\alpha \Pi)(x) }{\ell} \splimit 0,
\end{equation}
which means that
\begin{equation}
\lnorm{p'_2 p' (\alpha \Pi)(x) }{\ell} \splimit 0.
\end{equation}
Using the relation $p_2' p' \alpha = \alpha p_2 p$ together with properties
\ref{compatibilityProp} and \ref{limitProp}, and noting that 
$\alpha$ is invertible on $\alpha V_2$, it follows that
\begin{equation}
\lnorm{p_2 p \Pi(x) }{\ell} \splimit 0
\csq
\lnorm{\Pi(x) -  (p_0+p_1) p \Pi(x) }{\ell} \splimit 0;
\end{equation}
thus $V_0 \oplus V_1$ is sp-app for $\Pi$. Due to the minimality
of $\npspace{\Pi}$, this is only possible if $V_2 = \{0\}$; hence
$\alpha \npspace{\Pi} = \alpha V_1 = \npspace{\alpha \Pi}$.
\end{proof}

The properties \ref{compatibilityProp}--\ref{normProp} requested for $\alpha$
are fulfilled by a number of relevant transformations.

\emph{Lorentz transformations:} A Lorentz transformation $\alpha = \alpha(\Lambda)$
acts on $\cinftys^\ast$ ``as usual'' [i.e. through $\mathrm{ad}\, U(\Lambda)$],
on $\PhiFH^{\otimes n}$ in the same way on every tensor factor, and on $\mkr$ by
$\alpha .x = \Lambda x$, which is extended to $\mkr^n$ componentwise.
The properties \ref{limitProp} and \ref{normProp}
are obvious. Applying $\mathrm{ad}\, U(\Lambda)$ to the approximating
sequences in Eq.~\eqref{heuristProdDefn}, it is also easy to see that
$\alpha(\Pi(x)) = (\alpha \Pi)(\alpha.x)$. So Theorem~\ref{covarThm} tells us that
$\npspace{\alpha(\Lambda) \Pi}  = \alpha(\Lambda) \npspace{\Pi} $; the normal
product spaces are Lorentz covariant {\em as vector spaces}. 
Note that $\npspace{\Pi}$ is not necessarily stable under $\alpha(\Lambda)$,
since possibly $\Pi \neq \alpha(\Lambda)\Pi$; we would have to 
pass to a closure $\hat{\mathrm{N}} [\Pi] = \lspan \bigcup_\Lambda \npspace{\alpha(\Lambda)\Pi} $
if we aim at a decomposition of $\Lambda \mapsto \alpha(\Lambda)$
into irreducible subrepresentations.

\emph{Other symmetries} with ``geometric action,'' such as
dilations [$\alpha(\lambda).x = \lambda x$] and inner
symmetries ($\alpha.x = x$), show the same behavior as Lorentz transformations,
as long as they are unitary implemented
and fulfill certain regularity properties
(cf. Lemma~5.5 of Ref.~\onlinecite{Bos:Operatorprodukte}).
Since for our construction, it suffices to use a local unitary implementation
rather than a global one, it does not matter whether
the symmetries are broken or unbroken.\cite{BrokenSymm}

\emph{Derivatives:} To treat linear differential operators in our context,
it suffices to consider first order operators $D_\mu$, which act on
$\cinftys^\ast$ through $i \lbrack P_\mu,\cdotarg\rbrack$. Since they
leave $\PhiFH$ invariant, \cite{Bos:fields-article} they also act on
$\PhiFH^{\otimes n}$ by a formal product rule. As $D_\mu$ satisfies the
product rule on $\boundedops$, one may establish
\begin{equation}
  D_\mu \big( \Pi(x) \big) = \big(D_\mu \Pi \big)(x).
\end{equation}
[To see this, note that the approximating sequences ``$A_r \to \phi$''
can be chosen to be ``smeared'' with some test function $f_r$ --
compare the remark after the proof of Lemma~3.5 in Ref.~\onlinecite{Bos:fields-article} --
such that $D_\mu A_r = D_\mu(\hat A_r(f_r)) = \hat A_r (- \partial_\mu f_r)$,
so $D_\mu$ preserves the localization of the operator sequence.] Again,
property~\ref{normProp} is obvious, so the differential operators $D_\mu$
fulfill properties \ref{compatibilityProp} to \ref{normProp},
with $D_\mu$ acting trivially on $\mkr^n$. By concatenation and linear combination, 
the same is then true
for linear differential operators $D$ of arbitrary order. Hence we have
\begin{equation}
     \npspace{D\Pi} = D \npspace{\Pi}.
\end{equation}
The perturbative analog to this relation is known as
{\em Lowenstein's rule}. \cite{KelKop:composite_operators_1}

\section{Conclusions and outlook}        \label{conclsect}

In the course of the present paper, we have given a rigorous model-independent
proof of operator product expansions, based on a physically motivated assumption
that was formulated as a phase space condition.
In this context, product expansions are asymptotic series
in the short distance limit; their singular behavior is bounded by an inverse power.
We have introduced normal products in the sense of vector spaces that consist of
all fields contributing to the product expansion (up to a given level of accuracy).
These vector spaces show the expected properties, 
such as Lorentz covariance and Lowenstein's rule.

Originally, Wilson \cite{Wil:non-lagrangian} proposed operator product expansions
as a substitute for the Lagrangian, as a method of defining field theoretic models.
Though it would seem exaggerated to aim at constructive approaches from our results,
they might indeed serve as a basis for the classification of models. For example,
they could give a well-defined sense to the concept of local field equations:
The famous $\phi^4$ equation
\begin{equation}
  (\dalembert + m^2) \phi = \lambda \wickprod{\phi^3}\,, \;\;
  \lambda \neq 0,
\end{equation}
well known in perturbation theory,\cite{Zim:field_equations_phi4} can be introduced
in our context as
\begin{equation} \label{ourphi4}
(\dalembert + m^2) \phi  \in \npspace{\phi^{\otimes 3}} \; \backslash \; \cbb \phi.
\end{equation}
At present, it is unknown whether such a relation is stringent enough to define
a field theory (nor, in fact, whether it is compatible with any field theory at all).
There is strong evidence\cite{Fro:triviality} that in physical space-time,
the standard lattice approximation approach does not lead to a theory that fulfills
Eq.~\eqref{ourphi4}; however, other methods have been proposed that might
result in such a solution.\cite{Klauder} Equation \eqref{ourphi4} at least allows us to pose
the existence problem of $\phi^4_4$ independent of 
specific construction schemes.

More generally, it seems interesting to what extent field equations -- or
other properties of product expansions -- can define a field theory uniquely.
One encounters some obvious obstructions here, since there exist nontrivial
theories with a trivial field content\cite{TrivialFieldContent} 
$\PhiFH = \cbb \idop$, which might always occur as a tensor factor.
We can exclude these components, however,
by defining the following subnet $\afk_F$ of $\afk$ which may be
regarded as the ``point field part'' of the theory
(as remarked in Ref.~\onlinecite{Bos:fields-article}):
\begin{equation}
  \afk_F (\ocal) :=   \pcal(\ocal) '',
\end{equation}
where $\pcal(\ocal)$ is the polynomial algebra generated by all $\phi(f)$ with $\phi \in \PhiFH$,
$\supp f \subset \ocal$. 

In models which are generated by observable point fields 
(such as the free-field examples in Ref.~\onlinecite{Bos:fields-article}),
we have $\afk=\afk_F$,
and one would hope to find a description of $\afk$
in terms of field equations or similar relations.
In the presence of gauge fields, 
on the other hand, it might happen that $\afk_F \subsetneq \afk$,
since $\afk$ may include inherently non-point-like observables like Wilson loops
or Mandelstam strings. In this case, it is possible 
that the dynamics of the system cannot be described in terms of $\PhiFH$ alone,
but that field equations need to involve the extended objects mentioned.
Still, it would be worthwhile to ask what physical properties
(such as cross sections) are determined by $\afk_F$ only.
However, the details of such an analysis remain vague at the present stage. 

\begin{acknowledgments}
The author is indebted to D.~Buchholz for his support and for many helpful
hints and discussions.

The work has profited from
financial support by Evangelisches Studienwerk, Villigst,
which the author gratefully acknowledges.

\end{acknowledgments}

\input products-bbl.inc
%\bibliography{../qft}

\end{document}

%% file: products-macros.inc
\newtheorem{defn}{Definition}
\newtheorem{lemm}[defn]{Lemma}

\newtheorem{thm}[defn]{Theorem}

\DeclareMathOperator{\img}{img}
\DeclareMathOperator{\lspan}{span}
\DeclareMathOperator{\supp}{supp}

\newcommand{ \cinftys } {  \ccal^\infty(\Sigma) }
\newcommand{\lnorm}[2]{  \| #1 \|^{( #2 )}  }
\newcommand{\biglnorm}[2]{  \big\| #1  \big\|^{( #2 )}  }
\def \PhiFH { \Phi_{ \mathrm{FH} }  }

\def \cbb{\mathbb{C}}
\def \nbb{\mathbb{N}}
\def \rbb{\mathbb{R}}

\def \ccal {\mathcal{C}}

\def \hcal {\mathcal{H}}

\def \mcal {\mathcal{M}}

\def \ocal {\mathcal{O}}
\def \pcal {\mathcal{P}}

\def \scal {\mathcal{S}}

\def \vcal {\mathcal{V}}

\def \afk  {\mathfrak{A}}
\def \bfk  {\mathfrak{B}}

\def \pfk  {\mathfrak{P}}

\newcommand{\idop}{\openone}

\def \cinftys {  \ccal^\infty(\Sigma) }
\newcommand{\mkr}{\mcal}
\newcommand{\vlk}{ \vcal_+ }
\newcommand{\dist}{ \mathrm{dist}\, }
\newcommand{\evp} { ^\lbrack \!\,' \!\,^\rbrack }

\newcommand{\xmy}{ x \! - \! y }
\newcommand{\splimit}{ \xrightarrow[]{\mathrm{sp}} }
\newcommand{\wickprod}[1] {  : \!\! #1 \!\! : }
\newcommand{\npspace}[1] {  \mathrm{N} \lbrack #1 \rbrack }

\def \cdotarg { \, \cdot \, }

\def \im  {\mathrm{Im}\,}

\def \< {\langle}
\def \> {\rangle}

\def \dalembert { \pmb \Box }

\def \half {\frac{1}{2}}

\def \csq {\quad \Rightarrow \quad}

\def \equivalent {\quad \Leftrightarrow \quad}

\def \restrict {\lceil}

\newcommand{\hrskp}[2]{( #1 | #2 ) }

\def \boundedops {\bfk(\hcal)}

\def \tracenorm#1 { \| #1 \| _1 }
\def \hsnorm#1 { \| #1 \| _2 }